\documentclass[journal=jacsat,manuscript=article]{achemso}

\usepackage[version=3]{mhchem} % Formula subscripts using \ce{}

\usepackage{amsmath,amssymb}%Eq. and symbols
\usepackage{amsthm}%proof environment
\usepackage{tikz}%plot figure
\usepackage{braket}%Dirac bracket
\usepackage{xr-hyper}%cross reference, must be loaded before hyperref
\usepackage[colorlinks, linkcolor=red, anchorcolor=green, citecolor=blue]{hyperref}
\usepackage[ruled]{algorithm2e}%ruled makes caption at topside
\usepackage{multirow}%table multirow
\usepackage[normalem]{ulem}%delete line
\usepackage{booktabs}

\author{Yifan Ke}
\affiliation{Hefei National Research Center for Physical Sciences at the Microscale, and Hefei National Laboratory, University of Science and Technology of China, Hefei, Anhui 230026, China}
\author{Chuanjing Zeng}
\affiliation{Hefei National Research Center for Physical Sciences at the Microscale, and Hefei National Laboratory, University of Science and Technology of China, Hefei, Anhui 230026, China}
\author{Xinming Qin}
\affiliation{Hefei National Research Center for Physical Sciences at the Microscale, and Hefei National Laboratory, University of Science and Technology of China, Hefei, Anhui 230026, China}
\author{Wei-Lin Tu}
\affiliation{Graduate School of Science and Technology, Keio University, Yokohama, Kanagawa 223-8522, Japan}
\affiliation{Keio University Sustainable Quantum Artificial Intelligence Center (KSQAIC), Keio University, Tokyo 108-8345, Japan}
\author{Wei Hu}
\email{whuustc@ustc.edu.cn}
\affiliation{State Key Laboratory of Precision and Intelligent Chemistry, Department of Chemical Physics, and Anhui Center for Applied Mathematics, University of Science and Technology of China, Hefei, Anhui 230026, China}
\author{Jinlong Yang}
\email{jlyang@ustc.edu.cn}
\affiliation{State Key Laboratory of Precision and Intelligent Chemistry, Department of Chemical Physics, and Anhui Center for Applied Mathematics, University of Science and Technology of China, Hefei, Anhui 230026, China}

\abbreviations{IR,NMR,UV}

\title{Dynamic Moiré Potentials and Robust Wigner Crystallization in Large-Scale Twisted Transition Metal Dichalcogenides}
\begin{document}

\begin{abstract}
Understanding the dynamical evolution of large-scale moir\'e systems is crucial for connecting theoretical predictions with experimental observations.
Here we develop a machine-learning-based workflow, integrating DeePMD and DeepH frameworks with first-principles calculations, to efficiently investigate time-dependent structural and electronic responses in twisted bilayer transition metal dichalcogenides (TMDs) with experimentally relevant moir\'e supercells containing over 3000 atoms.
Using $\mathrm{WS_2}$ as a representative system, we show that low-temperature lattice vibrations and relaxation deepen the moir\'e potential wells, narrow the lowest conduction band, and facilitate the formation of strongly localized electronic states.
Based on DFT-derived moir\'e potentials that incorporate these dynamical effects, density-matrix-renormalization-group (DMRG) simulations reveal robust Wigner crystallization and a kagom\'e-patterned three-electron state, consistent with recent experimental observations.
Our workflow provides a practical route for exploring large moir\'e supercells beyond static configurations and offers new insight into the interplay between lattice dynamics, electronic localization, and emergent correlated states in twisted two-dimensional materials.
\end{abstract}

%\keywords{Wigner crystal, moir\'e material, machine learning, density functional theory, density matrix renormalization group}

\section{Introduction}

Wigner crystals represent a fundamentally important class of electron-ordered states that emerge when Coulomb repulsion dominates over kinetic energy in low-density systems~\cite{wigner1934interaction}. They provide a platform for exploring strongly correlated physics with potential implications in tunable electronic phases and emergent many-body phenomena. One conventional route to realizing Wigner crystallization is through Landau quantization in strong perpendicular magnetic fields, where the quenched kinetic energy promotes electron localization at fractional filling factors (e.g., $\nu=1/5$, $1/3$)~\cite{zhao2018crystallization,villegas2021competition,madathil2023moving}. However, magnetic-field-based approaches offer limited tunability of the underlying potential landscape, motivating the search for alternative systems that allow more flexible control over electron crystallization.

Twisted bilayer transition metal dichalcogenides (TMDs) have recently emerged as a powerful platform for realizing Wigner and generalized Wigner crystal states without magnetic fields~\cite{klein2001electronic,li2024wigner,huang2024electronic,tang2023evidence,padhi2021generalized,morales2023magnetism,matty2022melting,yannouleas2024wigner,xiang2025imaging}. In these moir\'e superlattices, carriers introduced via electrostatic doping become trapped in a triangular array of potential minima~\cite{li2024imaging}, where the moir\'e potential can exceed the exciton kinetic energy and generate extremely flat minibands~\cite{jin2019observation,xiong2023correlated}. As a result, TMD moir\'e homobilayers provide deep, highly tunable potential wells~\cite{erkensten2024stability,kiper2025confined,yannouleas2024crystal,dolgirev2024local} controlled by twist angle, strain, and carrier density~\cite{zhou2021bilayer,regan2020mott,yannouleas2023quantum,hu2025topological}. Despite these advantages, theoretical studies typically rely on simplified $\mathrm{\it{C}_{\rm{3}}}$-symmetric potential models~\cite{yannouleas2024wigner,reddy2023artificial,li2024wigner,li2025emergent}, and numerical methods including exact diagonalization and Hartree–Fock predict kagom\'e-like configurations at three-electron filling~\cite{ung2023competing,reddy2023artificial,pawlowski2024interacting,yannouleas2023quantum,zhou2024quantum}. Yet, first-principles modeling of realistic moir\'e supercells—often exceeding $10^3$ atoms at twist angles $\theta\le3^\circ$—remains computationally prohibitive~\cite{liu2025dpmoire,lucignano2019crucial,lisi2021observation,li2024evolution}. Furthermore, experimental systems exhibit thermal fluctuations, reconstruction, and disorder~\cite{ren2023impact,rosenberger2020twist}, which are not captured in static or symmetry-idealized theories, leaving significant gaps in understanding how dynamical lattice environments influence Wigner crystallization.

\begin{figure*}[!htb]
\centering
\includegraphics[width=1.0\linewidth]{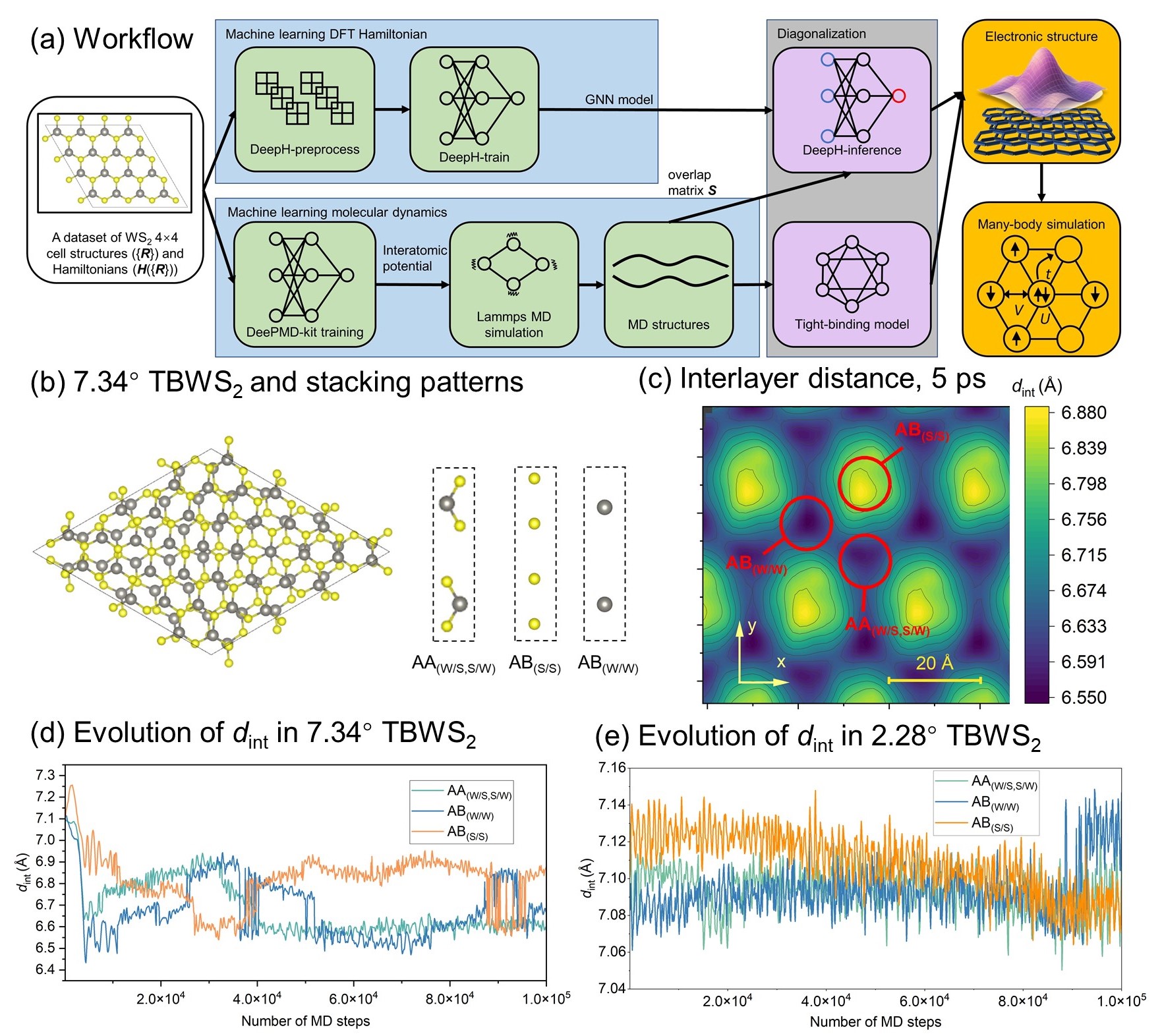}
\caption{(a) Schematic workflow of the moir\'e material molecular dynamics and electronic structure prediction. (b) A moir\'e unit cell of $\rm{TBWS_{2}}$. Different stacking area are present in the unit cell, including $\rm{AA_{(W/S,S/W)}}$, $\rm{AB_{(S/S)}}$, $\rm{AB_{(W/W)}}$. (c) Contour map of the $\rm{7.34^{\circ}}$ $\rm{TBWS_{2}}$ at the 10,000th molecular dynamics timestep (5 ps). (d) Interlayer distances ($\it{d}_{\rm{int}}$) of different stacking area of $\rm{7.34^{\circ}}$ $\rm{TBWS_{2}}$ with respect to the molecular dynamics steps. (e) $\it{d}_{\rm{int}}$ of $\rm{2.28^{\circ}}$ $\rm{TBWS_{2}}$ with respect to the molecular dynamics steps.}
\label{fig1}
\end{figure*}

Recent developments in machine-learning interatomic potentials and machine-learned Hamiltonians provide a promising route to overcome these limitations. Approaches such as DeePMD-kit~\cite{wang2018deepmd,zhang2018deep,zeng2023deepmd,batzner20223,nielsen2023accurate,jinnouchi2019fly} enable large-scale molecular dynamics with near-DFT accuracy, while DeepH-pack and its E(3)-equivariant extension~\cite{gong2023general,li2022deep} offer efficient mappings from atomic configurations to DFT-quality electronic Hamiltonians. These ML-based methods have been successfully applied to complex materials and moir\'e systems~\cite{liu2025dpmoire}, providing access to realistic structural fluctuations and electronic properties at a scale unattainable with conventional DFT.

In this work, we integrate DeePMD molecular dynamics with DeepH-based electronic-structure predictions to investigate twisted bilayer $\mathrm{WS_2}$ across experimentally relevant twist angles. Our simulations capture low-frequency interlayer “breathing” modes that dynamically reshape the moir\'e potential landscape and modulate carrier localization on picosecond timescales. We show that molecular-dynamics-relaxed structures deepen the moir\'e potential wells relative to rigid configurations, and density-matrix-renormalization-group (DMRG) calculations based on DFT-derived potentials reveal enhanced formation of kagom\'e-patterned Wigner crystals at triple-electron filling. This combined workflow enables a realistic, temperature-dependent description of structural and electronic responses in moir\'e TMDs and provides quantitative insight into how lattice dynamics influence correlated electronic phases.

\section{Results}\label{results}

\begin{figure*}[!htb]
\centering
\includegraphics[width=1.0\linewidth]{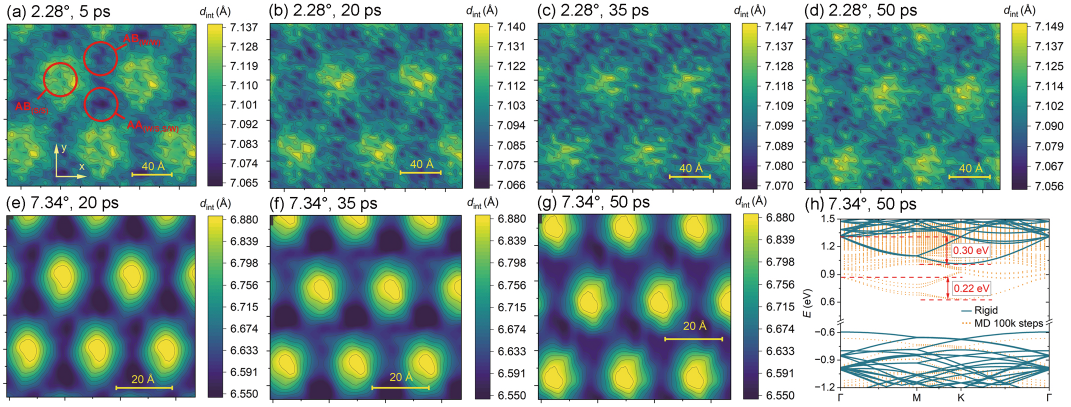}
\caption{(a)-(d) Contour maps of the interlayer distance $\it{d}_{\rm int}$ in $\rm{2.28^{\circ}}$ $\mathrm{TBWS_2}$ at different time frames. The snapshot frame is fixed, so the bright spots, indicating the largest interlayer distance, shift over time. (e)-(g) $\it{d}_{\rm int}$ contour maps of $\rm{7.34^{\circ}}$ $\mathrm{TBWS_2}$ at different time frames. (h) Electronic band structures of rigid and dynamic $\rm{2.28^{\circ}}$ $\mathrm{TBWS_2}$ structures.}
\label{fig2}
\end{figure*}

For the machine-learning molecular dynamics simulations, we employ the DeePMD-kit framework~\cite{wang2018deepmd,liu2025dpmoire,zeng2023deepmd}, while electronic structure calculations are performed using a combination of direct DFT via HONPAS~\cite{qin2015honpas}, machine-learned Hamiltonians via DeepH-pack~\cite{li2022deep}, and tight-binding model methods~\cite{long2022atomistic,nakhaee2020tight}.
The overall workflow is illustrated in Fig.~\ref{fig1}(a). We begin by constructing a comprehensive $\mathrm{WS_2}$ dataset, followed by machine learning procedures that significantly reduce computational cost. For the ML-based DFT Hamiltonian generation, we integrate DeepH-pack and its equivariant extension DeepH-E3 with HONPAS~\cite{gong2023general}, enabling efficient mapping from atomic structure to DFT-level Hamiltonians.

In the lower branch of the workflow, we use DeePMD-kit to train interatomic potential functions for $\mathrm{WS_2}$ and simulate molecular dynamics trajectories at small twist angles. Representative MD snapshots are then selected for further analysis: we compute conduction band structures using a tight-binding model that qualitatively reproduces the band flattening trend with decreasing twist angle, and also calculate overlap matrices for DeepH-based inference. Both the tight-binding and neural network approaches bypass the time-consuming self-consistent loops inherent in traditional DFT, providing an efficient route to explore electronic structures.
Furthermore, the obtained moiré electronic structures serve as the basis for constructing effective many-body Hamiltonians, enabling the investigation of strongly correlated phases such as Wigner crystals.

Using this pipeline, we analyze systems across a series of small twist angles and at various MD time frames, enabling both static and dynamical comparisons of electronic properties.
Fig.~\ref{fig1}(b) shows the twisted bilayer $\mathrm{WS_2}$ structure ($\rm{TBWS_2}$), where three highly symmetric stacking motifs—AA (W/S and S/W), AB (W/W), and AB (S/S)—emerge as key features of the moir\'e pattern.

\begin{figure*}[!htb]
\centering
\includegraphics[width=1.0\linewidth]{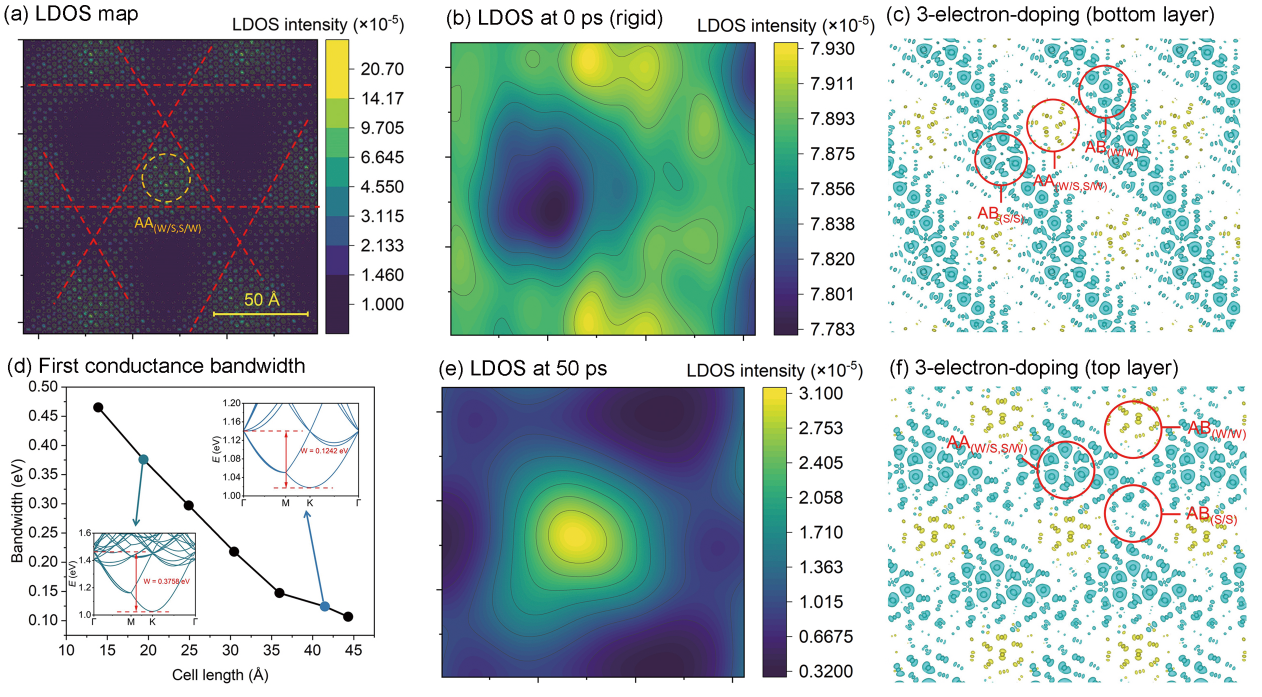}
\caption{(a) Local density of states map of the bottom layer of $\rm{2.28^{\circ}}$ $\rm{TBWS_2}$ at 50 ps. Red dotted lines outline the kagom\'e pattern formed by the bright regions, and the light yellow circle indicates the $\rm{AA_{(W/S,S/W)}}$ stacking area.
(b) Local density of states map of $\rm{2.28^{\circ}}$ $\rm{TBWS_2}$ at 0 ps, corresponding to a flat bilayer configuration.
(c) Effect of DFT+U on the localized states of the bottom layer of $\rm{2.28^{\circ}}$ $\rm{TBWS_2}$ at 0 ps, with $\it{U}$ = 3.0 eV applied to the W 5d orbitals.
(d) Bandwidth of the first conduction band as a function of the moir\'e supercell length for twist angles of $\rm{13.17^{\circ}}$, $\rm{9.43^{\circ}}$, $\rm{8.26^{\circ}}$, $\rm{7.34^{\circ}}$, $\rm{6.01^{\circ}}$, $\rm{5.09^{\circ}}$, and $\rm{4.41^{\circ}}$.
(e) Local density of states map of $\rm{2.28^{\circ}}$ $\rm{TBWS_2}$ at 50 ps, showing a more concentrated bright region compared with that in (b).
(f) DFT+U result for the top layer corresponding to the configuration in (c). Compared with the bottom layer in (c), the top layer exhibits a more pronounced kagom\'e-like pattern, attributed to the Coulomb interaction introduced by the DFT+U correction.
}
\label{fig3}
\end{figure*}

In Fig.~\ref{fig1}(c), we present a contour plot of the interlayer distance $d_{\mathrm{int}}$ for $\theta = 7.34^\circ$ $\mathrm{TBWS_2}$, where bright regions highlight local maxima in the moir\'e lattice.
Fig.~\ref{fig1}(d) and (e) show the time evolution of interlayer distances at the characteristic stacking sites. The relative ordering of these distances changes over time, indicating interlayer vibrations reminiscent of a ``breathing" mode. Notably, this breathing is more pronounced and frequent at $\theta = 7.34^\circ$ [Fig.~\ref{fig1}(d)], while at $\theta = 2.28^{\circ}$ [Fig.~\ref{fig1}(e)] the oscillations are smoother and slower—consistent with the larger moir\'e supercell and lower structural stiffness at small twist angles.

\subsection{Breathing interlayer distance}

The time-dependent variations in interlayer distance reflect the dynamic shifting of potential peaks and valleys, offering insight into the real-time evolution of the moir\'e potential landscape. While direct in situ detection of atomic configurations during experiments remains unfeasible, large-scale material simulations provide valuable indirect understanding of the structural dynamics in experimental samples.
Figs.~\ref{fig2}(a)–(d) show contour maps of the interlayer distance $d_{\mathrm{int}}$ for $\theta = 2.28^\circ$ $\mathrm{TBWS_2}$ at selected MD snapshots, while Figs.~\ref{fig2}(e)–(g) display analogous maps for $\theta = 7.34^\circ$. In each case, the spatial region is fixed for consistent comparison. The bright regions—corresponding to local maxima in interlayer spacing—clearly shift over time, evidencing the lattice’s breathing motion. Moreover, the amplitude of $d_{\mathrm{int}}$ fluctuations is approximately 0.1 Å for the $2.28^\circ$ system and about 0.3 Å for $7.34^\circ$, indicating that moir\'e superlattices with smaller twist angles exhibit more stable and inert breathing dynamics. This trend is consistent with their larger moir\'e unit cells and reduced local curvature, which collectively suppress structural fluctuations and enhance lattice rigidity.

In moir\'e materials, decreasing the twist angle results in increasingly flattened electronic bands. The reduced bandwidth enhances carrier localization within the enlarged moir\'e unit cells, promoting the formation of localized states~\cite{ding2025electric}.
We have fitted the electronic structure of $\mathrm{TBWS_2}$ using an 11-orbital tight-binding model and computed the conduction band structures across a series of twist angles ranging from $13.17^\circ$ to $2.28^\circ$ (see Supplementary Information).

In addition to the geometric effects of twist-angle reduction, thermal relaxation further promotes band flattening. 
As shown in Fig.~\ref{fig2}(h), a comparison between band structures from rigid and molecular-dynamics-relaxed geometries reveals a noticeable reduction in bandwidth when relaxation is included. This effect enhances the localization of carriers within the moir\'e potential, thereby facilitating the experimental realization of Wigner crystals. 
In the Supplementary Information, we compare the potential wells extracted from the MD-relaxed structures with those of the rigid configuration (Fig.~S12), showing that thermal relaxation leads to significantly deeper potential wells.

\subsection{Localized potentials}

The wave-like relaxation of twisted structures gives rise to a spatially varying moir\'e potential, determined by the underlying electronic structure. To probe this effect, we analyze the local density of states (LDOS) of the first conduction band and visualize its spatial distribution.
Fig.~\ref{fig3}(a) show LDOS contour map of $\theta = 2.28^\circ$ $\mathrm{TBWS_2}$ at the 100,000th MD step (50 ps, bottom layer). The moir\'e cell length, $\lambda_{\theta=2.28^{\circ}} \approx 80\,$ \AA, is about half the side length of the square simulation cell and yields seven periodic LDOS maxima. These bright regions mark areas of enhanced LDOS, suggesting potential localization centers for doped electrons in the first conduction band, which poses a kagom\'e pattern marked with red dotted lines.
Fig.~\ref{fig3}(b) and (e) present the LDOS distribution on the bottom layer, which comparatively shows the LDOS situation of rigid structure and the MD relaxed structure. It can be seen from the difference between the maximum and the minimum that MD relaxed structure has deeper and more triangular like LDOS potential well. 

To capture the qualitative effect of electron–electron interactions under doping, we employ the DFT+$U$ method in HONPAS, applying a Hubbard $U=3.0$ eV to the W 5d orbitals~\cite{csacsiouglu2011effective}—a commonly adopted value for 5d transition metals—to illustrate the on-site Coulomb–induced LDOS redistribution. Figs.~\ref{fig3}(c) and \ref{fig3}(f) show the LDOS difference between undoped and doped systems, with yellow and cyan isosurfaces indicating regions of increased and decreased LDOS, respectively. The on-site repulsion suppresses charge accumulation at atomic centers and redistributes the electronic density across the moir\'e unit cell. As seen in Fig.~\ref{fig3}(f), the doped system exhibits a lower maximum LDOS amplitude relative to the undoped case, reflecting the enhanced Coulomb repulsion that prevents excessive localization at the potential wells.

In the bottom layer, LDOS is enhanced near the $\rm{AA_{(W/S,S/W)}}$ region but reduced at both $\rm{AB_{(W/W)}}$ and $\rm{AB_{(S/S)}}$. Conversely, the top layer shows pronounced localization around the moir\'e $\rm{AB_{(W/W)}}$ region, accompanied by a triangular LDOS depletion at $\rm{AA_{(W/S,S/W)}}$. These contrasts indicate that electron doping drives charge redistribution, where on-site Coulomb repulsion expels electrons from $\rm{AA_{(W/S,S/W)}}$ potential wells and enhances LDOS in neighboring $\rm{AB_{(W/W)}}$ regions.

\begin{figure*}[!htb]
\centering
\includegraphics[width=1.0\linewidth]{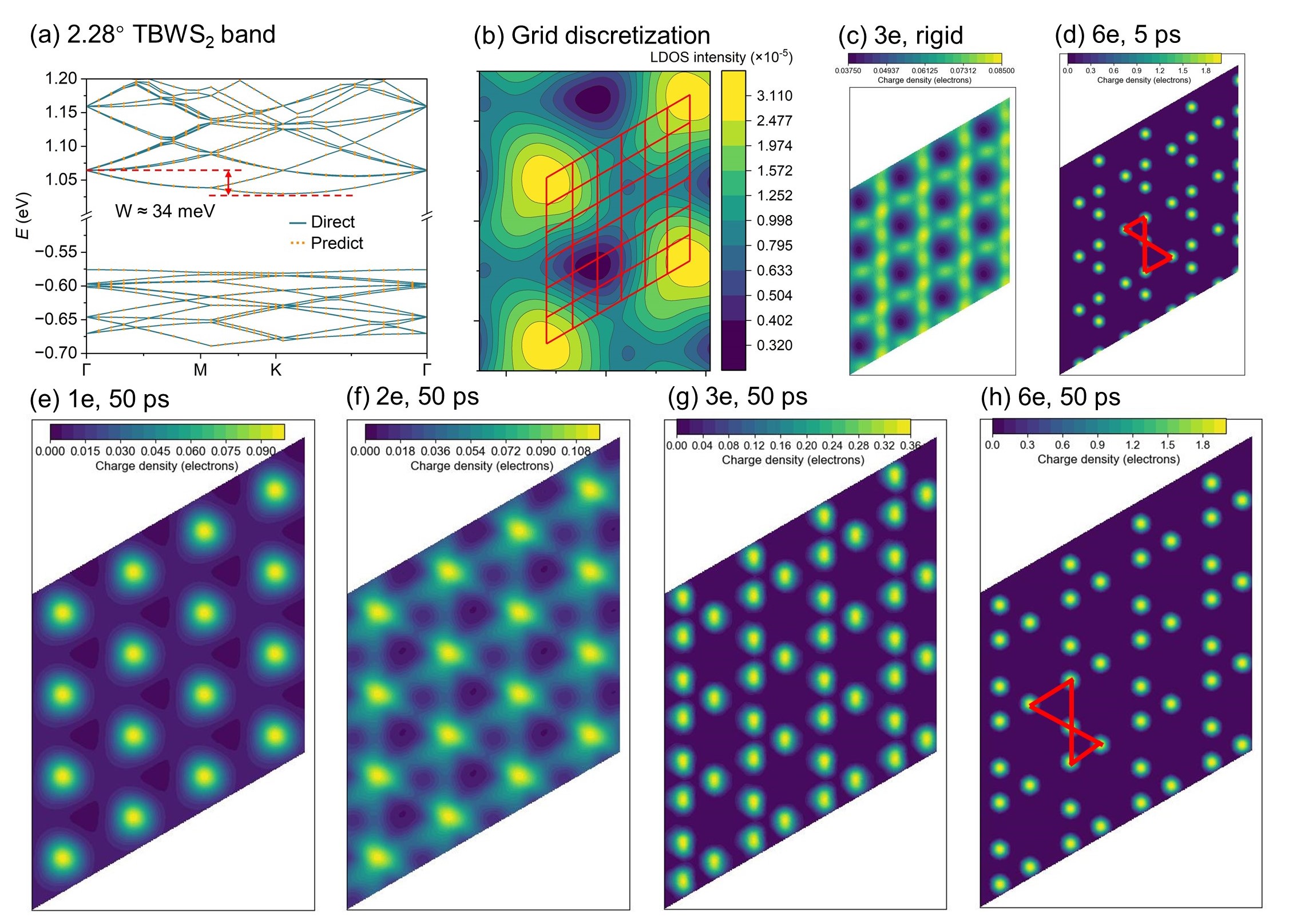}
\caption{(a) Electronic band structure of $\rm{2.28^{\circ}}$ $\rm{TBWS_2}$ at 50 ps, where solid lines denote direct DFT results and dots represent machine-learning predictions. (b) LDOS of the same structure in (a). The overlaid red grid marks the $\rm{7 \times 7}$ triangular lattice used in the tensor-network simulations, with the highlighted region corresponding to a single moir\'e cell. (c–h) Electron distributions from DMRG calculations at different fillings and timesteps: (c) three electrons of rigid structure, (d) six electrons at an early MD step (5 ps); (e) one, (f) two, (g) three, and (h) six electrons at a later stage (50 ps) per moir\'e cell. 
The deepest moir\'e potential wells coincide with the LDOS maxima in (b), pinning the electrons at the bright spots. Panels (c–g) show the $\rm{4 \times 4}$ extension of a single moir\'e cell, revealing periodic triangular-lattice patterns at 1e and 2e doping, and kagom\'e-lattice patterns at 3e and 6e doping. The corresponding dilute doping levels in the rescaled $\rm{7 \times 7}$ lattice are 1/98, 2/98, 3/98, and so on.}
\label{fig4}
\end{figure*}

This redistribution trend is further consistent with the twist-angle–dependent evolution of the conduction bandwidth: as shown in Fig.~\ref{fig3}(d), the first conduction band narrows systematically as the moir\'e lattice constant increases, decreasing from 0.3758 eV at $9.43^\circ$ to 0.1242 eV at $4.41^\circ$. Since the moir\'e lattice constant grows with decreasing twist angle, smaller angles naturally yield narrower bandwidths and stronger localization tendencies. In line with this progression, the $2.28^\circ$ structure—whose bandwidth (0.034 eV) is obtained from the DFT calculation in Fig.~\ref{fig4}(a)—exhibits the narrowest conduction band, signifying the strongest electronic confinement and the most favorable conditions for Wigner crystal formation in the largest moir\'e supercells.

\subsection{Model calculations}

We performed model Hamiltonian calculations based on moir\'e potential wells fitted from LDOS data. Instead of employing the $\it{C}_{\rm{3}}$-symmetric potential, our approach directly extracts the moir\'e potential profile from the LDOS distribution of the dynamical $\rm{2.28^{\circ}}$ $\rm{TBWS_2}$ structure at 50 ps, and employs it to perform DMRG calculations on a triangular lattice discretizing a single moir\'e unit cell.
The reliability of this method is further supported by recent DMRG studies on extended Hubbard models of moir\'e superlattices~\cite{biborski2025charge,kumar2025origin}, which have demonstrated its accuracy in capturing generalized Wigner crystal phases.
A $\rm{7\times7}$ rhombohedral patch of the triangular lattice is then adopted to encompass the potential-well region [Fig.~\ref{fig4}(b)]. 
The site-dependent potential term $\Delta_i$ in Eq.~\ref{Hubbard_Hamiltonian} is obtained from the LDOS triangular map, while the hopping energy is estimated from the bandwidth of the first conduction band [Fig.~\ref{fig4}(a)]. The Hubbard $\it{U}$ and Coulomb $\it{V}$ are estimated as 30.0 and 25.0, respectively, with further details provided in Method~\ref{effective_hamiltonian}.

When the system is doped with one electron per moir\'e unit cell, the electrons become pinned at the potential wells coinciding with the LDOS maxima, collectively forming a triangular lattice, as shown in Fig.~\ref{fig4}(e). At a higher doping level of three electrons per moir\'e cell, the carriers localize within the potential wells but also experience strong mutual Coulomb repulsion. Consequently, they arrange into a triangular configuration consistent with the underlying $\it{C}_{\rm{3}}$-symmetric potential landscape, giving rise to a moir\'e electron crystal with an equilateral triangular pattern. 
Due to lattice relaxation, similar to that illustrated in Fig.~\ref{fig3}(a), the resulting electron molecules deviate from a perfect equilateral geometry and instead adopt a distorted triangular arrangement.

In the 2e-doping case, the additional electron forms a density distribution extended along the triangular lattice lines [Fig.~\ref{fig4}(f)], exhibiting in-plane $x$–$y$ symmetry without favoring any of the three moir\'e vertices. This differs from Hartree–Fock results~\cite{li2024wigner}, which typically show two bright spots, and likely reflects the stronger correlation effects captured by DMRG.
At 3e filling, the rigid structure displays a more diffuse charge distribution with reduced peak intensity [Fig.~\ref{fig4}(c)], whereas the MD-relaxed configuration [Fig.~\ref{fig4}(d)] shows sharper localization arising from relaxation-induced deepening of the moir\'e potential wells. These structural differences provide the foundation for the higher-filling patterns observed at larger doping.
As the system evolves from 3e to 6e doping, electrons populate moir\'e sites in the order indicated in Fig.~\ref{fig4}(g), with intermediate configurations provided in the Supplementary Information. At 6e doping, a pronounced kagom\'e lattice emerges [Fig.~\ref{fig4}(h)], forming a well-defined triangular network centered on the moir\'e sites, in agreement with both theoretical predictions and experimental observations~\cite{li2024wigner,reddy2023artificial}.
Furthermore, the comparison between Figs.~\ref{fig4}(d) and \ref{fig4}(h) demonstrates that stacking-dependent interlayer breathing modulates the depth and orientation of the potential wells, thereby shaping the resulting localization patterns and determining the geometry of the kagom\'e lattice. The small red triangles in (d) and (h) highlight how the kagom\'e orientation evolves over time under thermal breathing effects.
Together, these results clarify how low-temperature moir\'e breathing dynamics govern the filling sequence and spatial arrangement of Wigner-crystal states, suggesting new avenues for controlling electron crystallization through structural relaxation.

\section{Summary}\label{discussion}

In summary, we develop an integrated workflow that combines machine-learning molecular dynamics with a machine-learned Hamiltonian to capture the temperature-dependent structural evolution and electronic responses of twisted bilayer $\mathrm{WS_2}$. Our simulations reveal low-frequency breathing modes that modulate interlayer separations and dynamically reshape the moir\'e potential wells, leading to bandwidth narrowing and enhanced LDOS localization. Using DFT-derived potentials that incorporate these thermal fluctuations, DMRG calculations demonstrate robust Wigner crystallization, including a kagom\'e-patterned three-electron state, with MD-relaxed structures exhibiting stronger localization than rigid configurations. These results highlight twisted TMD moir\'e materials as a highly tunable platform for exploring strongly correlated electron phases on a triangular lattice~\cite{wang2020correlated,devakul2021magic,shabani2021deep,tilak2022moire} and provide an efficient, scalable route for future studies of moir\'e-driven electronic crystals.

\section{Methods}\label{method}

\subsection{Machine learning workflow}

We prepared the dataset for DeePMD training by performing ab initio molecular dynamics (AIMD) simulations of bilayer $\rm{WS_{2}}$ at an initial temperature of 100 K using the ABACUS package~\cite{zeng2023deepmd,wang2018deepmd,li2016large,lin2024ab}. A total of 500 time steps were simulated, and the resulting molecular dynamics (MD) frames were randomly divided into training and validation datasets. The DeePMD model was trained for 10 million epochs, achieving a final energy mean absolute error (MAE) of 1.03 meV and root mean square error (RMSE) of 1.22 meV. The corresponding force MAE and RMSE were 2.14 meV/\AA and 2.70 meV/\AA, respectively. MD predictions were subsequently carried out using the trained interatomic potential within the LAMMPS interface~\cite{thompson2022lammps}. Several representative frames were extracted for further electronic structure calculations using HONPAS, DeepH+HONPAS, and the tight-binding method. The DeepH+HONPAS code is publicly available on GitHub~\cite{ke2025combining,datasetYifan}.

\subsection{Effective Hamiltonian model}\label{effective_hamiltonian}
We have adopted the Hubbard model with moir\'e potential wells. To simulate the electron distribution on the lattice, we have focused on a smaller area of the overall triangular lattice, a rhombic area divided into a $\rm{7\times 7}$ lattice. The moir\'e potential is applied to the lattice sites by evaluating the potential well function at different coordinates. Besides, the electron repulsions between nearest sites are in the form of Coulomb interaction~\cite{kim2024dynamical}. The Hamiltonian can be constructed as:
\begin{equation}
\begin{aligned}
H = &\sum_{\langle ij\rangle,\sigma} t\, c_{i,\sigma}^{\dagger} c_{j,\sigma} 
   + \sum_{i,\sigma} \Delta_{i} c_{i,\sigma}^{\dagger} c_{i,\sigma} \\
&+ \sum_{i} U\, c_{i,\uparrow}^{\dagger} c_{i,\uparrow} 
             c_{i,\downarrow}^{\dagger} c_{i,\downarrow} 
   + \sum_{i,j,\sigma} V(r)\, c_{i,\sigma}^{\dagger} c_{i,\sigma}\, c_{j,\sigma}^{\dagger} c_{j,\sigma}.
\end{aligned}
\label{Hubbard_Hamiltonian}
\end{equation}
In our simulations, the nearest hopping parameter $\it{t}$ is fixed to 1.0. The on-site interaction $\it{U}$ is estimated as the Coulomb energy within a moir\'e unit cell, $\it{U}\,\rm{\sim}\, e^{2}/(4\pi\epsilon_0\epsilon \it{a}\rm{)}$, where $\it{a}$ is the effective lattice spacing of the discretized triangular mesh. Similarly, the nearest-neighbor interaction is taken as $\it{V}\,\rm{\sim}\, e^{2}/(4\pi\epsilon_0\epsilon \it{r}_{NN}\rm{)}$, with $\it{r_{NN}}$ the distance between neighboring sites. For $\rm{2.28^{\circ}}$ $\rm{TBWS_2}$, $a$ and $r_{NN}$ are near 10 \AA ($\lambda_{\theta=2.28^{\circ}}\approx 80$ \AA), and $\epsilon \approx 10$ ($\rm{WS_2}$ 8$\sim$12).
The hopping amplitude is approximated as $\it{t}\sim W\rm{/9}$, where $\it{W}$ (Fig.~\ref{fig4}(a)) denotes the conduction bandwidth and the factor 9 accounts for the triangular discretization of the moir\'e unit cell. Although these Coulomb estimates yield comparable magnitudes for $\it{U}$ and $\it{V}$, the stronger orbital overlap within a Wannier state implies $\it{U} > \it{V}$. Accordingly, we set $\it{U}\,\rm{=}\, 30.0$ and $\it{V}_{\rm{0}}\,\rm{=}\,25.0$ in our calculations.
The Coulomb interaction is included with a cutoff radius six times the nearest-neighbor lattice spacing, which effectively accounts for almost all relevant Coulomb interactions.
We have also tested $\it{U}$ values from 20 to 40, finding no significant change in the Wigner crystal, indicating that it remains stable throughout this parameter range.

The site-dependent potential term $\Delta_i$ is extracted from the DFT LDOS map. To ensure that the extracted potential is on the same energy scale as $\it{t}$ and $\it{U}$ while preserving the shape of the triangular wells, the LDOS data are rescaled by a factor of $10^{5}$. The intersite Coulomb interaction is modeled as $\it{V}_{\rm{0}}\rm{/}\it{r}$ with $\it{V}_{\rm{0}}\,\rm{\approx}\,25.0$. The doped electrons are initialized in an antiferromagnetic configuration. DMRG simulations are performed using the TenPy package~\cite{tenpy2024}, benchmarked on small chain and square-lattice systems before being applied to the moir\'e potential well case. The bond dimension is set to 2000, the maximum truncation error to $10^{-4}$, and the maximum number of sweeps to 10,000. These parameters are chosen to reproduce the Wigner crystal features observed experimentally~\cite{li2024wigner}. Additional electron distributions obtained under different parameter settings are provided in the Supplementary Information.

\section{Acknowledgments}

Y. K. and W.-L. T. appreciate the fruitful discussion with Yusuke Nomura and Shunsuke Furukawa.
This work is partly supported by the Innovation Program for Quantum Science and Technology (2021ZD0303306), the Strategic Priority Research Program of the Chinese Academy of Sciences (XDB1170000, XDB0450101), the National Natural Science Foundation of China (22503093, 22288201, 22173093, 21688102), and the National Key Research and Development Program of China (2016YFA0200604, 2021YFB0300600). W.-L. T. was supported by the Center of Innovation for Sustainable Quantum AI (JST Grant Number JPMJPF2221) and JSPS KAKENHI Grant Number JP25H01545. The AI-driven experiments, simulations and model training were performed on the robotic AI-Scientist platform of Chinese Academy of Sciences. The authors thank the Hefei Advanced Computing Center, the Supercomputing Center of Chinese Academy of Sciences, the ORISE Supercomputing Center, the Supercomputing Center of USTC, the National Supercomputing Centers in Wuxi, Tianjin, Shanghai, and Guangzhou for the computational resources. The authors acknowledge the Beijing Beilong Super Cloud Computing Co., Ltd for providing HPC resources (http://www.blsc.cn/).

\bibliography{reference}% common bib file

\end{document}

\title{Supplementary Information - Dynamic Moiré Potentials and Robust Wigner Crystallization in Large-Scale Twisted Transition Metal Dichalcogenides}

\author{Yifan Ke}
\affiliation{Hefei National Research Center for Physical Sciences at the Microscale, and Hefei National Laboratory, University of Science and Technology of China, Hefei, Anhui 230026, China}

\author{Chuanjing Zeng}
\affiliation{Hefei National Research Center for Physical Sciences at the Microscale, and Hefei National Laboratory, University of Science and Technology of China, Hefei, Anhui 230026, China}

\author{Xinming Qin}
\affiliation{Hefei National Research Center for Physical Sciences at the Microscale, and Hefei National Laboratory, University of Science and Technology of China, Hefei, Anhui 230026, China}

\author{Wei-Lin Tu}
\affiliation{Graduate School of Science and Technology, Keio University, 3-14-1 Hiyoshi, Kohoku, Yokohama, Kanagawa 223-8522, Japan}
\affiliation{Keio University Sustainable Quantum Artificial Intelligence Center (KSQAIC), Keio University, Tokyo 108-8345, Japan}

\author{Wei Hu}
\email{whuustc@ustc.edu.cn}
\affiliation{State Key Laboratory of Precision and Intelligent Chemistry, Department of Chemical Physics, and Anhui Center for Applied Mathematics, University of Science and Technology of China, Hefei, Anhui 230026, China}

\author{Jinlong Yang}
\email{jlyang@ustc.edu.cn}
\affiliation{State Key Laboratory of Precision and Intelligent Chemistry, Department of Chemical Physics, and Anhui Center for Applied Mathematics, University of Science and Technology of China, Hefei, Anhui 230026, China}

\maketitle

\section{DFT calculation details}\label{calculation details}
\begin{figure*}[t]
\centering
\includegraphics[width=1.0\linewidth]{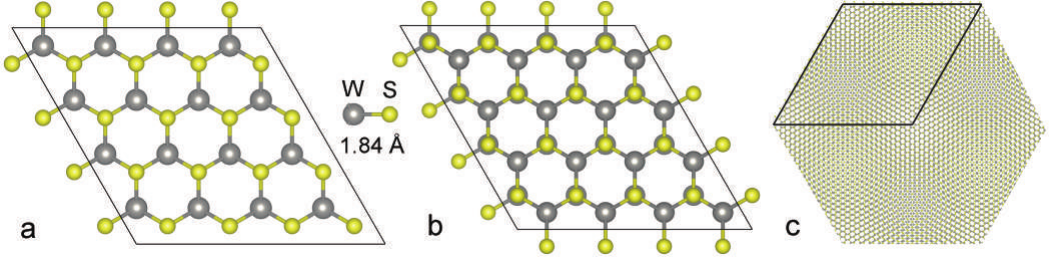}
\caption{(a) The bottom layer of bilayer $\rm{WS_2}$. (b) The top layer of bilayer $\rm{WS_2}$. (c) The unit cell of moir\'e $\rm{2.28^{\circ}}$ $\rm{TBWS_2}$.}
\label{single cell}
\end{figure*}

The crystal structure of the single-layered tungsten sulfide ($\rm{WS_{2}}$) material in this work is shown in Fig.~\ref{single cell}(a), and the bilayer $\rm{WS_{2}}$ is constructed by coping the bottom layer and shift in-plane so that the $\rm{W}$ atom of the top layer overlaps the $\rm{S}$ atom of the bottom layer (AB stacking), as shown in Fig.~\ref{single cell}(b). The single cell is expanded to a larger $4\times 4$ cell before performing lattice relaxation. The relaxation is done in Vienna Ab initio Simulation Package (VASP) with a K point sampling on a $2 \times 2 \times 1$ Monkhorst-Pack grid. The relaxed structures of both $\rm{WS_2}$ and $\rm{BWS_2}$ are using as the basic structures of all the following calculations. The relaxed interlayer distance of $\rm{BWS_2}$ (W element plane distance) is 7.01 \AA. Due to the computational limit of the small angle twisting case in plane-wave DFT package, the VASP calculations of $4\times 4$ cells serve as the benchmark for HONPAS, which uses atomic orbital basis and can efficiently handle large structures with thousands of atoms. The electronic band structures of the same structure but different DFT packages are compared. For HONPAS the effects of basis size is also considered, but this effect is minor to the localized LDOS and ESP phenomenon in large scale two-dimensional materials. The twisted bilayer $\rm{WS_{2}}$ ($\rm{TBWS_{2}}$) structure is constructed by setting the average interlayer distance obtained from the bilayer case, and transforming the bottom and top layer single cell respectively with a pair of commensurate rotation vectors that reflects the closest angle in experiment, which is near $\rm{2.0^{\circ}}$. We choose (15, 14) as the rotation vector setting, where the twist angle between two vectors $\vec{V_1}=15\,\vec{a_1}+14\,\vec{a_2}$ and $\vec{V_2}=14\,\vec{a_1}+15\,\vec{a_2}$ is $\rm{2.28^{\circ}}$, where $\vec{a_1}$ and $\vec{a_2}$ are cell vectors with the length of 3.19 \AA. The $\rm{2.28^{\circ}}$ $\rm{TBWS_{2}}$ moir\'e cell is shown in Fig.~\ref{single cell}(c).

\begin{figure*}[t]
\centering
\includegraphics[width=1.0\linewidth]{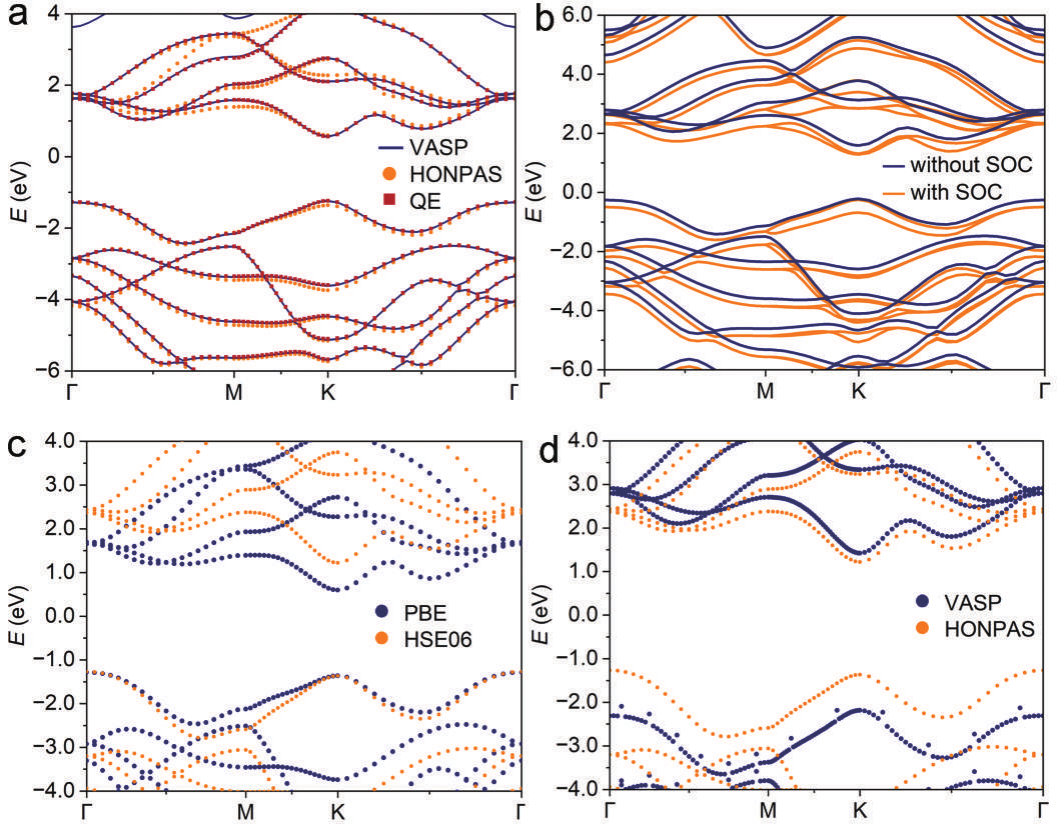}
\caption{(a) Electronic band structures of single layer and single cell $\rm{WS_{2}}$ calculated with PBE functional of different DFT packages: VASP, HONPAS, and Quantum-Espresso. (b) Spin-orbit-coupling (SOC) effects on the band structures of single layer and single cell $\rm{WS_{2}}$. This calculation is performed with VASP. SOC effects cause band splitting. (c) Functional effects on the band structure of single layer and single cell $\rm{WS_{2}}$. The PBE and HSE06 functional results are produced with HONPAS. The HSE06 functional leads to a larger band gap without affecting the top valence band. (d) Comparision between HSE06 results of single layer and single cell $\rm{WS_{2}}$ produced with VASP and HONPAS. They show similar band structures while VASP produces a larger band gap.}
\label{band_WS2_single}
\end{figure*}

We have also performed single cell calculations in VASP and HONPAS. The Monkhorst-Pack grid is set to $12 \times 12 \times 1$ for both packages and for HONPAS calculations the basis size is chosen to be DZP. The single cell is a single layer minimum cell of $\rm{WS_2}$ with the shape of X-M-X sandwich structure.
The electronic band structures calculated with VASP and HONPAS are presented in Fig.~\ref{band_WS2_single}. In HONPAS calculations, the basis size is set to be double-zeta polarized (DZP), and the K point sampling for SCF is also $2 \times 2 \times 1$ Monkhorst-Pack grid. The band gap of single layer $\rm{WS_2}$ is 1.65 eV in HONPAS and 1.77 eV in VASP, which is consistent with the DFT results of 1.6 eV in Ref.~\cite{klein2001electronic}. The band structure calculated with Quantum-Espresso is also consistent with VASP and HONPAS results. In Fig.~\ref{band_WS2_single}(b), we present the spin-orbit coupling effects in $\rm{WS_2}$, which causes the shift of band energies and band splitting. In Fig.~\ref{band_WS2_single}(c), we compare the band structures of PBE and HSE06, calculated with HONPAS. The PBE and HSE06 results share the same level of valence band maximum, but HSE06 accounts for the larger band gap (2.50 eV) than PBE (1.77 eV). In Fig.~\ref{band_WS2_single}(d), the HSE06 band structures calculated with VASP and HONPAS are compared. They share similar conductance band structures but shifted valence bands. However, despite the band gap differnce, VASP and HONPAS have consistent HSE06 band shapes of $\rm{WS_2}$.

\begin{figure*}[t]
\centering
\includegraphics[width=1.0\linewidth]{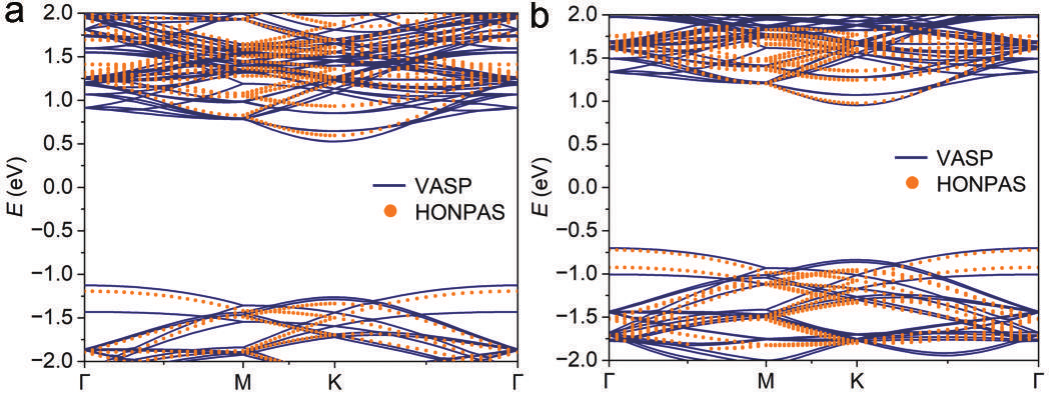}
\caption{Electronic band structures of (a) single layer $\rm{4 \times 4}$ cell $\rm{WS_{2}}$ ($\rm{WS_{2}}$ $\rm{4 \times 4}$) and (b) bilayer layer $\rm{4 \times 4}$ cell $\rm{WS_{2}}$ ($\rm{BWS_{2}}$ $\rm{4 \times 4}$) (AB stacking). These are PBE results. Due to band folding, they show different band structures from the single cell. VASP and HONPAS share consistent band structures.}
\label{band_WS2_BWS2_4_4_PBE}
\end{figure*}

\begin{figure*}[t]
\centering
\includegraphics[width=1.0\linewidth]{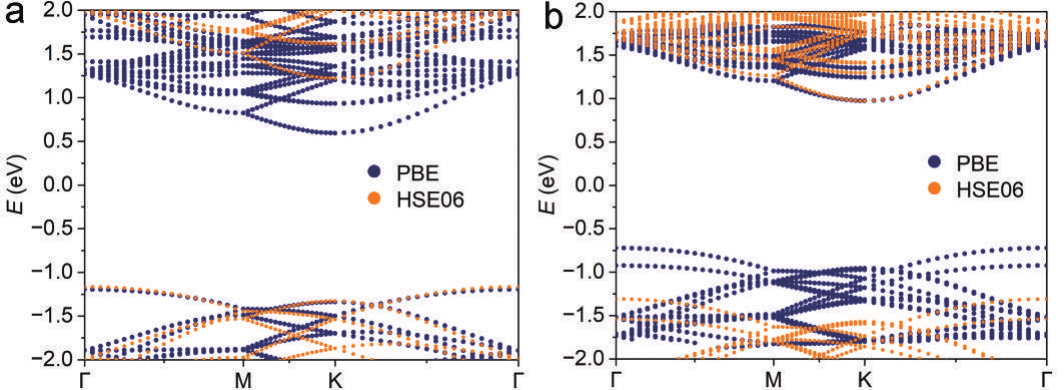}
\caption{Electronic band structures of (a) $\rm{WS_{2}}$ $\rm{4 \times 4}$ and (b) $\rm{BWS_{2}}$ $\rm{4 \times 4}$. These are HSE06 results calculated with HONPAS. For both structures, HSE06 produces larger band gaps.}
\label{band_WS2_BWS2_4_4_HSE06}
\end{figure*}

In Fig.~\ref{band_WS2_BWS2_4_4_PBE}, we present the PBE band structure comparison between VASP and HONPAS, showing good consistency. This agreement validates the reliability of HONPAS for subsequent calculations. Building on this, we further employ the HSE06 functional, which is widely regarded as a proper solution to the band gap underestimation problem in DFT. In Fig.~\ref{band_WS2_BWS2_4_4_HSE06}, we compare the HSE06 and PBE band structures of $\rm{WS_{2}}$ $\rm{4 \times 4}$ and $\rm{BWS_{2}}$ $\rm{4 \times 4}$. HSE06 produces larger band gaps, while the valence and conductance band shapes are the same as in PBE.

\begin{figure}[t]
\centering
\includegraphics[width=1.0\linewidth]{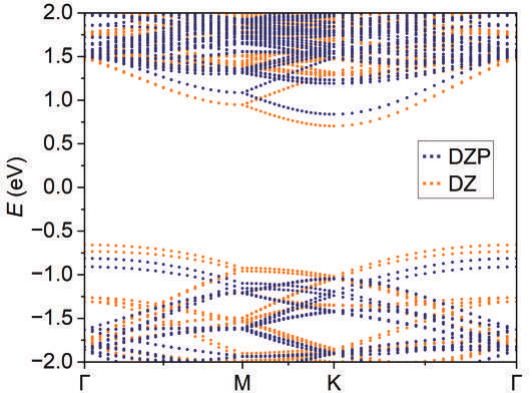}
\caption{Basis size effect on the electronic band structures. The system is $\rm{BWS_{2}}$ $\rm{4 \times 4}$ using PBE functional. In DZ basis, there are 12 orbitals for the $\rm{W}$ element and 8 orbitals for the $\rm{S}$ element. In DZP basis, there are 15 orbitals for the $\rm{W}$ element and 13 orbitals for the $\rm{S}$ element. DZP band gap is 0.29 eV larger.}
\label{band_DZ_DZP}
\end{figure}

We present the basis size effect in Fig.~\ref{band_DZ_DZP}. DZP basis uses more orbitals to describe an atom, and is accurate enough for the benchmark with plane-wave DFT package. In larger moir\'e cells, DZ basis is also useful when DZP basis is not computationally applicable. For example, in $\rm{BWS_{2}}$ $\rm{4 \times 4}$, the basis size effect has an influence of 0.29 eV band gap shifting, which is minor compared to the band gap (1.65 eV).

\begin{figure*}[t]
\centering
\includegraphics[width=1.0\linewidth]{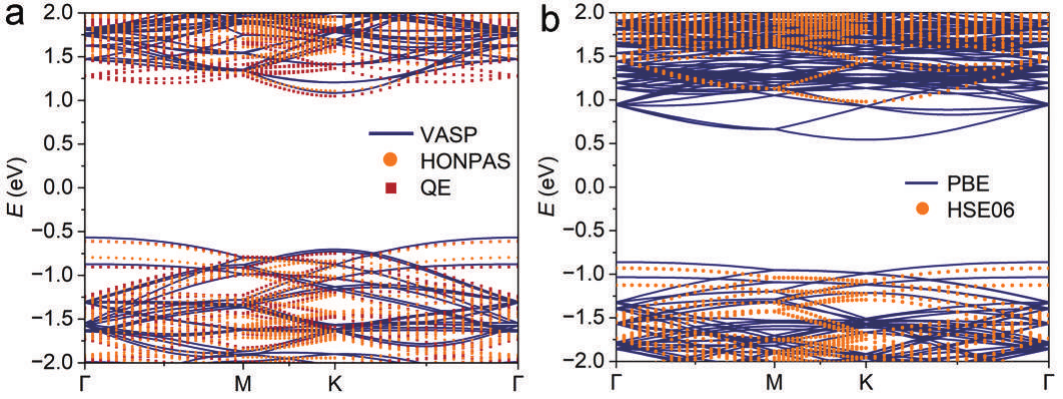}
\caption{(a) Electronic band structures of twisted bilayer $\rm{WS_{2}}$ with the twist angle of $\rm{13.17^{\circ}}$ ($\rm{13.17^{\circ}}$ $\rm{TBWS_{2}}$). VASP, HONPAS and Quantum-Espresso results are all in PBE functional. This moir\'e system contains 38 $\rm{W}$ and 76 $\rm{S}$ atoms. All results produce similar CBM and VBM, though in QE there is a deviation at the $\Gamma$ point. (b) PBE and HSE06 functional band structures of $\rm{9.43^{\circ}}$ $\rm{TBWS_{2}}$ calculated with HONPAS. HSE06 band gap is 0.51 eV larger. This moir\'e system contains 74 $\rm{W}$ and 148 $\rm{S}$ atoms.}
\label{band_TBWS2_13.17_9.43}
\end{figure*}

For $\rm{13.17^{\circ}}$ $\rm{TBWS_{2}}$, the band structure calculation is possible in plane-wave DFT packages. In Fig.~\ref{band_TBWS2_13.17_9.43}(a), PBE band structures of plane-wave DFT and atomic-orbital DFT are consistent. For a smaller twist angle of a larger moir\'e unit cell, such as $\rm{9.43^{\circ}}$ $\rm{TBWS_{2}}$, HONPAS calculation is computationally possible on a moderate cluster with 24 cores but not VASP or Quantum-Espresso. This indicates a superiority of atomic-orbital DFT in handling the moir\'e systems. In Fig.~\ref{band_TBWS2_13.17_9.43}(b), the HSE06 band gap is 0.51 eV larger than PBE.

\section{Tight binding model}\label{tb}
We adopt the 11-band tight binding model for fitting $\rm{WS_{2}}$, as this model is popular and useful for various TMD materials~\cite{peng2024modified,hu2021realistic,dias2018band}. The hopping energies can be formulated by the two-center approximation developed by J. C. Slater and G. F. Koster~\cite{slater1954simplified}. In the two-center approximation scheme, the hopping energies are orbital- and directional-dependent, with a reduced number of parameters. We have selected $\it{p_{x}}$, $\it{p_{y}}$, $\it{p_{z}}$ for two S atoms and $\it{d_{xy}}$, $\it{d_{yz}}$, $\it{d_{zx}}$, $\it{d_{x^{2}-y^{2}}}$, and $\it{d_{3z^{2}-r^{2}}}$ for the W atom. There are 7 paramters for Slater-Koster hoppings and 8 paramters for on-site energies. To get the best shot for the first conduct band fitting, we have used TBStudio, a package for tight-binding paramter fitting~\cite{nakhaee2020tight}. We have included the nearest W-S, W-W, and S-S hoppings in the model. The optimized parameters can be found in the Tab.~\ref{fitting}.

\begin{table*}[t]
    \centering
    \caption{Tight-binding model parameters for fitting $\mathrm{WS}_{2}$.}
    \ltx@label{fitting}
    \renewcommand{\arraystretch}{1.2}
    \setlength{\tabcolsep}{6pt}
    
    \begin{ruledtabular}
    \begin{tabular}{lcccccccc}
        \textbf{Type} & $d_{xy}$ & $d_{yz}$ & $d_{xz}$ & $d_{3z^{2}-r^{2}}$ & $d_{x^{2}-y^{2}}$ & $p_{x}$ & $p_{y}$ & $p_{z}$ \\
        \hline % 使用标准 hline，ruledtabular 会处理成漂亮的长横线
        \textbf{Energy (eV)} & -0.327 & -2.817 & -1.119 & -0.445 & -0.163 & -6.956 & -7.746 & -13.731 \\
    \end{tabular}
    \end{ruledtabular}
    
    \vspace{0.4cm}
    
    \begin{ruledtabular}
    \begin{tabular}{lcccccccc}
        \textbf{Type} & $pd\sigma$ & $pd\pi$ & $dd\sigma$ & $dd\pi$ & $dd\delta$ & $pp\sigma$ & $pp\pi$ & \\
        \hline
        \textbf{Energy (eV)} & -5.552 & 1.581 & -1.017 & 0.712 & 0.330 & 0.894 & -1.594 & \\
    \end{tabular}
    \end{ruledtabular}
\end{table*}

\begin{figure*}[t]
\centering
\includegraphics[width=1.0\linewidth]{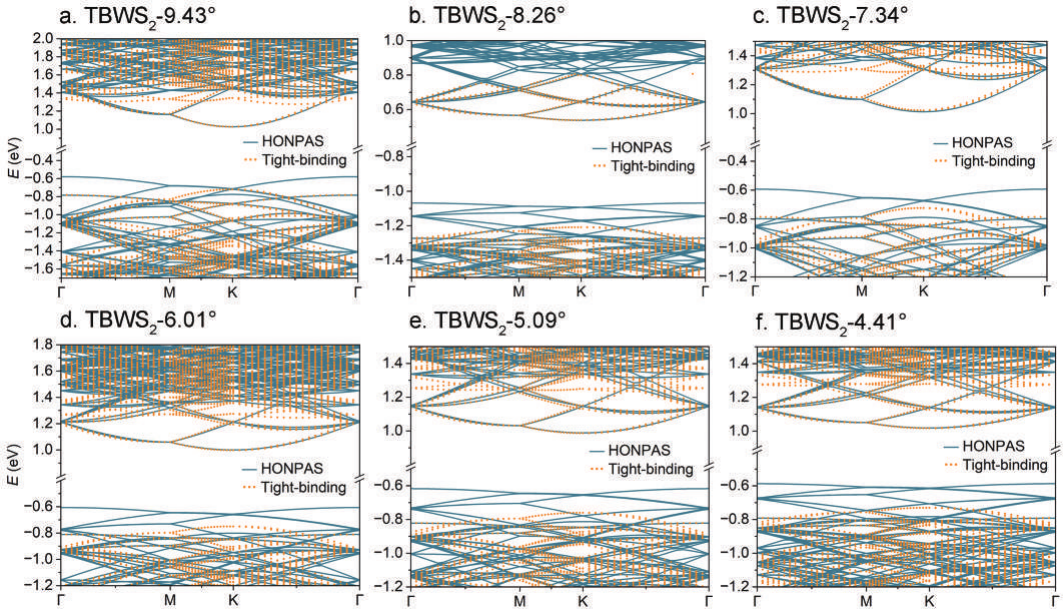}
\caption{Comparative electronic band structures of different twist angles between tight-binding model and HONPAS. From (a) to (f), 6 types of angles are considered, including $\rm{9.43^{\circ}}$, $\rm{8.26^{\circ}}$, $\rm{7.34^{\circ}}$, $\rm{6.01^{\circ}}$, $\rm{5.09^{\circ}}$, $\rm{4.41^{\circ}}$. }
\label{band_honpas_tb}
\end{figure*}

The fitted model has the good ability of computing the band structures for large twisted systems. In Fig.~\ref{band_honpas_tb}, a wide range of twist angles are considered, and the first conductance bands of all angles are aligned well with the DFT results. As the twist angle decreases, the band width of the first conductance band decreases, which is well captured by the tight-binding model.

\section{DeepH training}\label{deeph}

\begin{figure*}[t]
\centering
\includegraphics[width=0.5\linewidth]{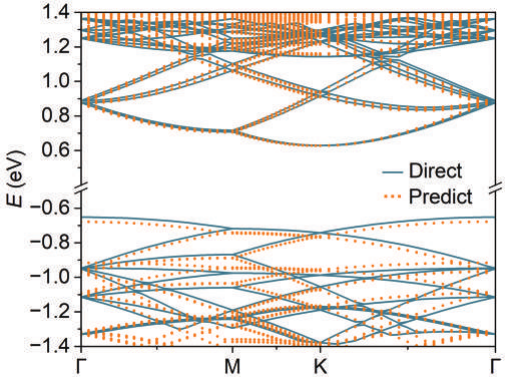}
\caption{DeepH predicted electronic band structure of the $\rm{8\times 8}$ supercell of $\rm{BWS_2}$.}
\label{band_BWS2_8_8_honpas_deeph}
\end{figure*}

The maximum matrix element in the training loss matrix is $\rm{3.3\times 10^{-5}\,eV}$. We have performed DeepH prediction on the trained model with an example of $\rm{8\times 8}$ supercell of $\rm{WS_2}$. The predicted band structure is good in Fig.~\ref{band_BWS2_8_8_honpas_deeph}. This system has 384 atoms, calculated with HONPAS PBE functional in DZ basis sizes and $\rm{4 \times 4 \times 1}$ k-grid Monkhorst Pack. In the NAO2GTO scheme in HONPAS, the NAO orbitals are fitted with Gaussian functions, which for $\rm{WS_2}$ material the maximum orbital radius is 7.36 \AA. To maintain consistency throughout the entire machine learning workflow—including preprocessing, training, and inference—we employ the same fitted GTO orbitals both in all HONPAS calculations and in the preparation of the DeepH dataset.

\section{Additional figures}\label{figures}
In this section, we present several supplementary figures of MD and relaxed structures of $\rm{TBWS_{2}}$. These include the interlayer distance distribution, the local density of states (LDOS) of the first conduction band, and the effects of doping.

\begin{figure*}[t]
\centering
\includegraphics[width=0.8\linewidth]{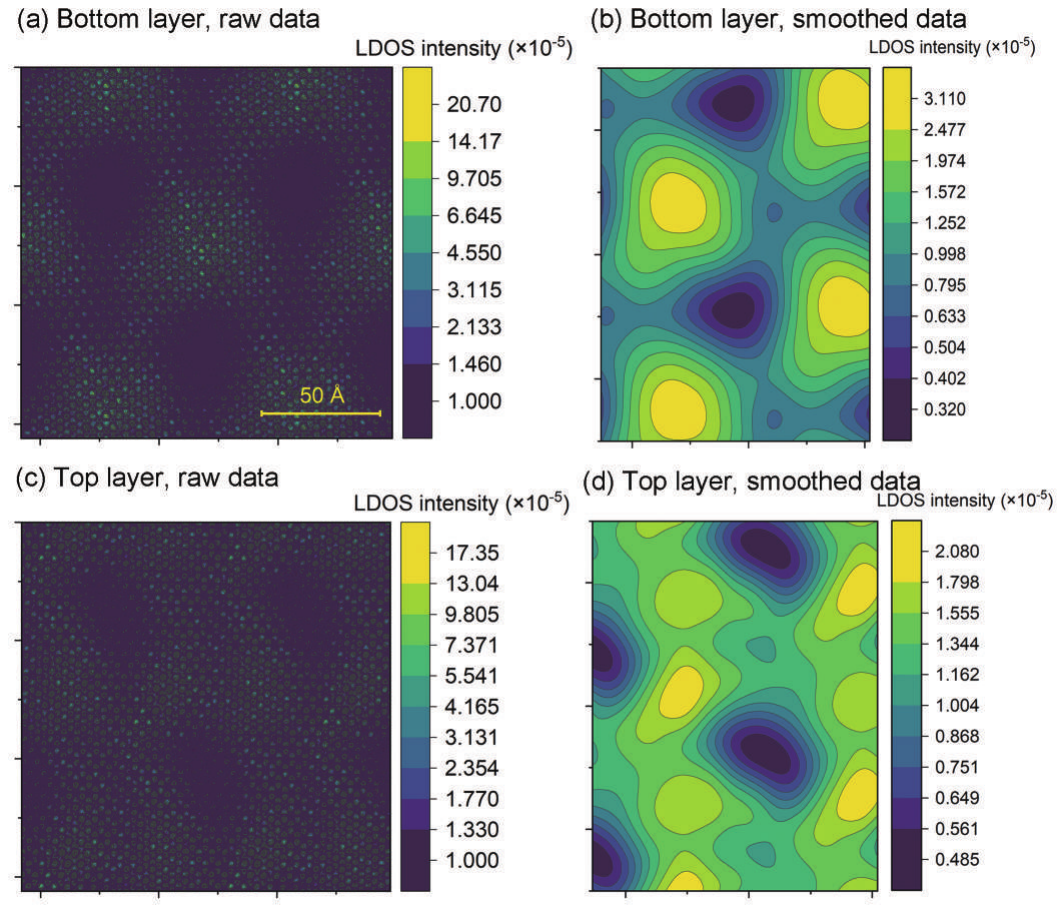}
\caption{(a) Raw bottom-layer LDOS of $\rm{2.28^{\circ}}$ $\rm{TBWS_{2}}$.
(b) Smoothed LDOS corresponding to (a).
(c) Raw top-layer LDOS of $\rm{2.28^{\circ}}$ $\rm{TBWS_{2}}$.
(d) Smoothed LDOS corresponding to (c).}
\label{LDOS_TBWS2_2.28_10W_0e}
\end{figure*}

\begin{figure*}[t]
\centering
\includegraphics[width=0.5\linewidth]{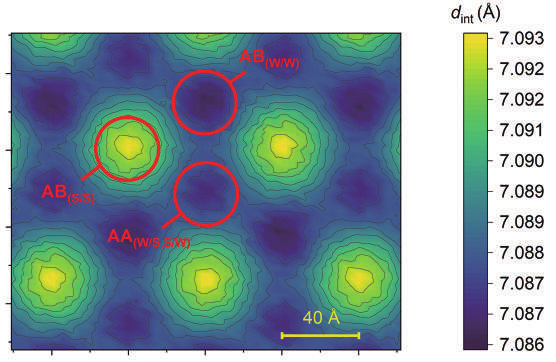}
\caption{Interlayer distance map of relaxed $\rm{2.28^{\circ}}$ $\rm{TBWS_2}$ obtained with HONPAS. The calculation employed a DZP basis set and a force tolerance of $\rm{0.2\,eV/}$ \AA.}
\label{dint_TBWS2_228_honpas_relaxed}
\end{figure*}

\begin{figure*}[t]
\centering
\includegraphics[width=1.0\linewidth]{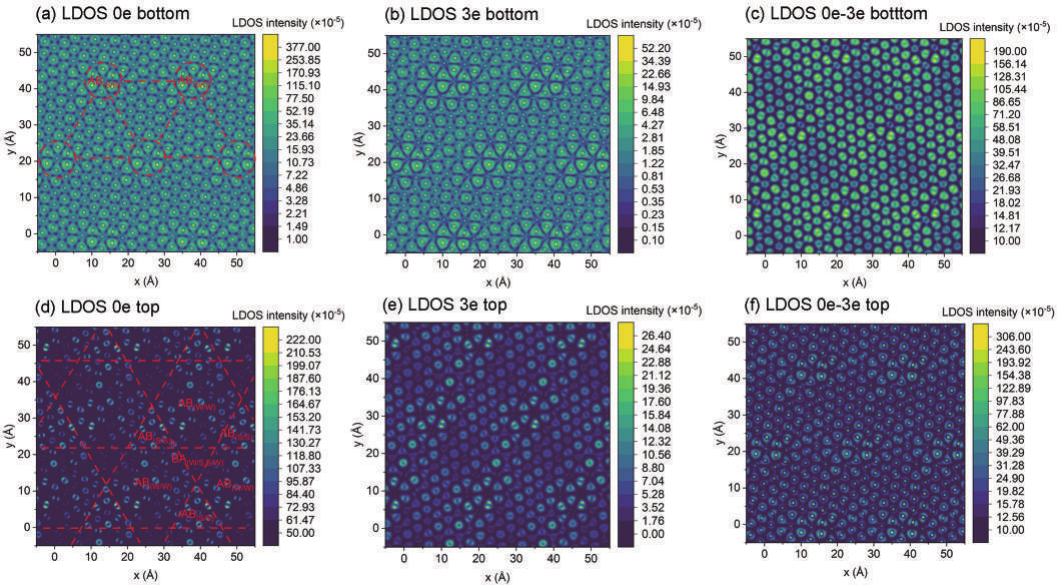}
\caption{(a) Local density of states (LDOS) of $\rm{7.34^{\circ}}$ $\rm{TBWS_{2}}$ at the 10,000 MD step. The bottom layer slice is cut at the same height of the W atoms, and the stacking patterns are labeled within the red dotted circles. (b) LDOS of the bottom layer, with the consideration of extra net charge of three electrons. (c) Differential LDOS map between undoped and three-electron-doped cases. (d)-(f) LDOS of the top layer. In (d), a structure of kagom\'e pattern is marked with red dotted lines, where the highlighted LDOS spots form the potential wells that introduce the crystallized array of electrons.}
\label{LDOS_TBWS2_7.34_1W}
\end{figure*}

\begin{figure*}[t]
\centering
\includegraphics[width=1.0\linewidth]{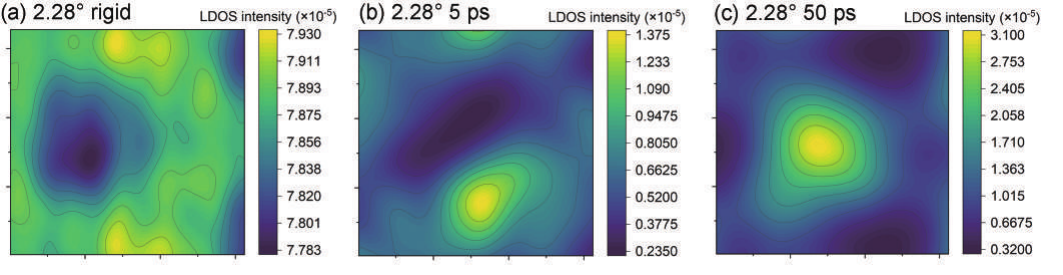}
\caption{Smoothed LDOS maps of $\rm{2.28^{\circ}}$ $\rm{TBWS_2}$ at different MD steps: (a) rigid structure, (b) 5 ps, and (c) 50 ps. Each snapshot covers a single moir\'e cell, with the potential well depth represented by the difference between the maximum and minimum values on the color bar.}
\label{raw_potential_compare}
\end{figure*}

\begin{figure*}[t]
\centering
\includegraphics[width=1.0\linewidth]{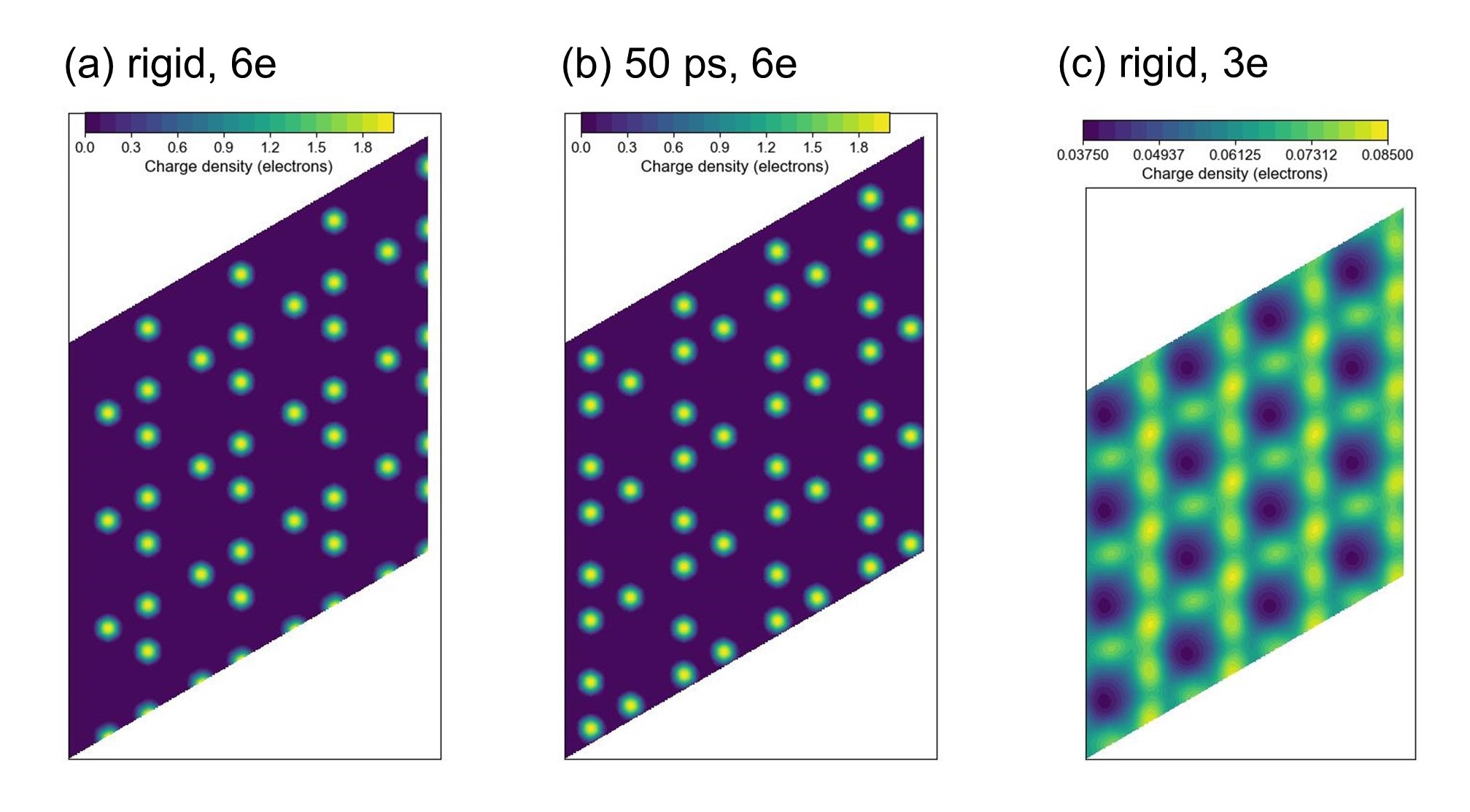}
\caption{(a) Six-electron-doped rigid $\rm{2.28^{\circ}}$ $\rm{TBWS_2}$ structure.
(b) Six-electron-doped $\rm{2.28^{\circ}}$ $\rm{TBWS_2}$ structure after 50 ps of MD relaxation.
(c) Three-electron-doped rigid structure. The charge density becomes more diffuse and delocalized, indicating a shallower potential well and a more fragile Wigner crystal pattern.}
\label{kagome_0W_10W}
\end{figure*}

\begin{figure*}[t]
\centering
\includegraphics[width=1.0\linewidth]{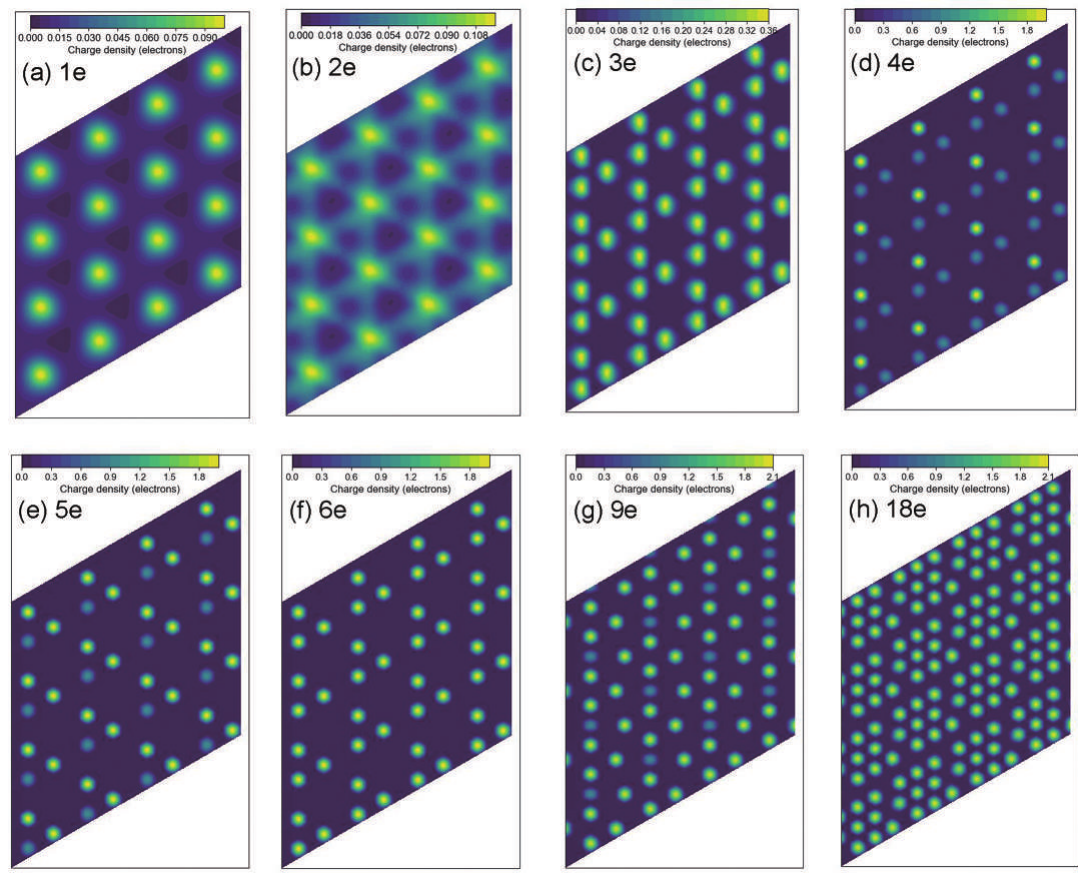}
\caption{DMRG results of electron distribution at different filling levels. (a)-(h) 1, 2, 3, 4, 5, 6, 9, 18 electrons respectively. The considered lattice is the $\rm{7\times 7}$ triangular lattice on the 50 ps time frame of the dynamical $\rm{2.28^{\circ}}$ $\rm{TBWS_2}$.}
\label{DMRG_all}
\end{figure*}

\begin{figure*}[t]
\centering
\includegraphics[width=1.0\linewidth]{3e_UV_compare.jpg}
\caption{Three-electron-doping cases with various Hubbard $U$ and Coulomb $V$ parameters. (a) $U$ = 20, $V$ = 15, (b) $U$ = 25, $V$ = 20, (c) $U$ = 35, $V$ = 20, (d) $U$ = 35, $V$ = 30, (e) $U$ = 40, $V$ = 25, (f) $U$ = 40, $V$ = 35, (g) $U$ = 50, $V$ = 25. The emergent electron crystal remains stable across a broad range of parameter configurations. We find that at three-electron doping, the Wigner crystal exhibits a kagom\'e-lattice pattern when $V$ $<$ $U$, and this structure persists as $U$ is varied from 20 to 50, indicating its robustness against changes in interaction strength.}
\label{3e_UV_compare}
\end{figure*}

\bibliographystyle{unsrt}
\bibliography{iopart-num}